\documentclass[a4paper,fleqn]{cas-sc}

\usepackage[numbers]{natbib}
\usepackage{amsmath,amsfonts}
\usepackage{amsfonts,amssymb}
\usepackage{algorithmic}
\usepackage[ruled,linesnumbered]{algorithm2e}
\usepackage{array}
\usepackage{xcolor}
\usepackage{textcomp}
\usepackage{stfloats}
\usepackage{url}
\usepackage{verbatim}
\usepackage{graphicx}
\usepackage{bm}
\usepackage{float}
\usepackage{subfigure}

\newcommand{\enc}[1]{\ensuremath{\mathsf{Enc}\left(#1\right)}}
\newcommand{\dec}[1]{\ensuremath{\mathsf{Dec}\left(#1\right)}}

\def\tsc#1{\csdef{#1}{\textsc{\lowercase{#1}}\xspace}}
\tsc{WGM}
\tsc{QE}
\tsc{EP}
\tsc{PMS}
\tsc{BEC}
\tsc{DE}

\newdefinition{rmk}{Remark}

\begin{document}
\let\WriteBookmarks\relax
\def\floatpagepagefraction{1}
\def\textpagefraction{.001}


\shortauthors{Z. Feng et~al.}

\title [mode = title]{Privacy-aware Fully Model-Free Event-triggered Cloud-based HVAC Control}                      


%
\author[1]{Zhenan Feng}[style=chinese]






\affiliation[1]{organization={Department of Electrical Engineering, City University of Hong Kong},
    addressline={Tat Chee Avenue, Kowloon}, 
    city={Hong Kong},
    }

\author[1]{Ehsan Nekouei}[style=chinese]
\cormark[1]
\ead{enekouei@cityu.edu.hk}
\cortext[cor1]{Corresponding author}


\nonumnote{The work was supported by the Research Grants Council of Hong Kong under Project CityU 21208921 and a grant from Chow Sang Sang Group Research Fund sponsored by Chow Sang Sang Holdings International Limited.}

\begin{abstract}
Privacy is a major concern when computing-as-a-service (CaaS) platforms, \emph{e.g.}, cloud-computing platforms, are utilized for building automation, as CaaS platforms can infer sensitive information, such as occupancy, using the sensor measurements of a building. Although the existing encrypted model-based control algorithms can ensure the security and privacy of sensor measurements, they are highly complex to implement and require high computational resources, which result in a high cost of using CaaS platforms. To address these issues, in this paper, we propose an encrypted fully model-free event-triggered cloud-based HVAC control framework that ensures the privacy of occupancy information and minimizes the communication and computation overhead associated with encrypted HVAC control. To this end, we first develop a model-free controller for regulating indoor temperature and CO$_2$ levels. We then design a model-free event-triggering unit which reduces the communication and computation costs of encrypted HVAC control using an optimal triggering policy.
Finally, we evaluate the performance of the proposed encrypted fully model-free event-triggered cloud-based HVAC control framework using the TRNSYS simulator, comparing it to an encrypted model-based event-triggered control framework, which uses model predictive control to regulate the indoor climate. Our numerical results demonstrate that, compared to the encrypted model-based method, the proposed fully model-free framework improves the control performance while reducing the communication and computation costs. More specifically, it reduces the communication between the system and the CaaS platform by 64\% amount, and its computation time is 75\% less than that of the model-based control.
\end{abstract}




\maketitle

\section{Introduction}
\subsection{Motivation}
As the requirements for precise process control and quality improvement become more advanced and complex, the necessity for designing suitable cloud-based controllers rises steadily. Cloud-based platforms are suitable for control scenarios because they can provide the required computational resources for systems with insufficient local computing resources. Utilizing insights from physical and mathematical principles of building systems, well-established model-based cloud control methods have been effectively implemented in building automation systems \cite{bird2022real,ZHOU2022104571}. Cloud-based control is applied to Heating, Ventilation, and Air Conditioning (HVAC) systems in buildings to provide thermal comfort and maintain indoor air quality. In a cloud-based HVAC control architecture, the sensor measurements from the system are sent to a cloud computing unit where the controller is hosted. This cloud-based controller calculates the control input based on the received measurements and then transmits the control input back to the system. There are two major issues that must be addressed in cloud-based HVAC control: (i) privacy and (ii) the cost of using computing-as-a-service platforms that offer cloud services.

Privacy is a primary concern for users of cloud-based control systems. Cloud controllers, although potentially trustworthy, could act as inquisitive adversaries, attempting to deduce confidential information from the system’s sensor data. Sensor measurements of a building are highly sensitive as building automation systems often rely on them to estimate occupancy status to determine whether a building is occupied. External service providers, such as cloud-based HVAC controllers, can potentially analyze these sensory data to infer occupancy-related details, raising significant privacy concerns for individuals within the building. For example, as demonstrated in \cite{co2estimation}, a neural network-based occupancy estimator using CO$_2$ data can achieve over $90\%$ predictive accuracy, indicating the potential risk of privacy breaches.

To address privacy concerns, Homomorphic Encryption (HE) technique has been developed to enhance data security in cloud computing by enabling operations on encrypted data. HE allows additions and multiplications to be performed directly on ciphertexts, enabling a cloud controller to receive encrypted data and output encrypted control inputs without needing decryption. This ensures that sensitive information remains confidential, preventing the cloud controller from extracting any details from the encrypted measurements.

However, computing-as-a-service platforms, \emph{e.g.}, cloud service providers, charge clients based on the amount of computing resources utilized \cite{ibm}. Although, HE ensures privacy and security, it significantly increases the required computing resources. Encrypted implementation of model-based control methods using HE results in two main challenges. First, model-based HVAC control methods, such as Model Predictive Control (MPC), involve repeated encrypted computations over multiple iterations. For instance, the computation time for unencrypted MPC and encrypted MPC with five iterations is $0.02$ and $5.74$ seconds, respectively, \emph{i.e.}, the computation time is $250$ times more under encryption, which implies a significant increase in the utilization of computational resources. 

Second, solving control problems with constraints in encrypted model-based control is challenging because it often requires multiple rounds of communication, where the control input needs to be decrypted and encrypted repeatedly. Encrypted model-free control offers a more efficient solution for encrypted cloud-based control by reducing communication and computation costs and eliminating the need for multiple communications between the client and the cloud.

Model-based approaches also require extensive model knowledge and associated processes. As HVAC systems grow in scale, the complexity of formulating thermal dynamics increases, making it challenging to meet rising expectations for control performance. Model-free control methods present an efficient alternative, eliminating the need for complex modeling tasks. These methods determine the dynamic behavior of the system using only input-output data, thereby simplifying implementation.

While encrypted model-free cloud-based control effectively maintains privacy and reduces computation costs, it has several limitations. Firstly, HE supports only additive and multiplicative operations, with many fully homomorphic encryption schemes limiting the number of multiplications. Secondly, existing encrypted model-free methods require the computation of certain nonlinear functions, such as the Sigmoid function, which is not supported by most HE schemes. Lastly, during the training and evaluation of the model-free controller, sensitive information could potentially be leaked, necessitating measures to preserve privacy throughout these processes.

\subsection{Contributions}
In this paper, we develop an encrypted model-free event-triggered cloud-based HVAC control system using the Homomorphic Encryption (HE) technique. Our goal is to enhance the efficiency of cloud-based control systems while ensuring data privacy. The main contributions of this work are summarized as follows: 
\begin{enumerate}[a)]
    \item An encrypted model-free cloud-based HVAC control method is developed. In this method, the system dynamics are modeled without relying on thermal dynamic equations, and the HVAC sensor measurements are encrypted using the HE scheme. This is the first attempt to combine HE technique in the model-free cloud-based HVAC control.  
    \item A model-free event-triggering unit is designed, where event decisions are made using neural networks instead of relying on predefined thresholds. This approach allows for more flexible and adaptive triggering behavior. Meanwhile, it reduces the communication and computation burdens in the encrypted cloud-based control system by reducing the frequency of communication between the HVAC system and the cloud significantly and, at the same time, maintains a desirable control performance.
    \item Unlike model-based control, the proposed encrypted fully model-free event-triggered cloud-based HVAC control framework can model the dynamic behavior of the unknown system only using data from the system’s inputs and outputs, without the need for mathematical models for implementation.
\end{enumerate}

\subsection{Related Work}
Homomorphic Encryption (HE) is an encryption method that permits computations directly on encrypted data without requiring decryption \cite{marcolla2022survey,acar2018survey}. HE schemes can be divided into two categories: Partially Homomorphic Encryption (PHE) and Fully Homomorphic Encryption (FHE). PHE only allows addition or multiplication to be performed on the ciphertexts. Examples of PHE schemes include multiplicative homomorphic encryption schemes represented by the RSA scheme \cite{RSA} and ElGamal scheme \cite{elgamal1985public} and additive homomorphic encryption schemes represented by the Paillier scheme \cite{paillier1999public}. Compared to PHE schemes, FHE schemes support both addition and multiplication to be performed on the ciphertexts. The CKKS scheme \cite{cheon2017homomorphic} is one of the most efficient fully HE schemes, which is commonly used in privacy-preserving machine learning \cite{ckksml1,ckksml2,ckksml3}.
 
With the development of the HE technique, it has been widely employed in secure control applications \cite{r2}. Encrypted model-based control has been studied in \cite{feng2023privacypreserving,lin2018secure,darup2017towards,r4}.  In \cite{r4}, the authors introduced a linear control scheme that employs public key encryption based on ElGamal homomorphic encryption \cite{elgamal1985public}. The stability analysis for a linear system governed by a dynamic ElGamal-encrypted system was investigated in \cite{Stability-ElGamal}. Lin \textit{et al.} \cite{lin2018secure} presented a secure controller for nonlinear systems and investigated the stability of its closed-loop performance. However, these approaches rely on the exact model of the system.

Model-based controllers may not always perform well, \emph{e.g.}, when unmodeled dynamic is present or when the exact model of the system is unknown.\cite{r11,safaei2018adaptive}. In such cases, model-free control methods are more desirable.
Encrypted model-free controllers have been studied in the literature, \emph{e.g.}, Encrypted State-Action-Reward-State-Action (encrypted SARSA) in \cite{r12,r13,r14} and encrypted Deep Q-Network (encrypted DQN) in \cite{r15}. The paper \cite{r12} proposed an encrypted version of SARSA based on the Paillier encryption scheme. However, encrypted SARSA and most encrypted conventional Reinforcement Learning (RL) methods require storing the Q-table, which limits their application to problems with large state and action spaces. To solve this problem, encrypted deep reinforcement learning has been proposed \cite{r15,suh2021encrypted}. Encrypted DQN uses neural networks to approximate the action-value function so that it can represent infinite Q-values.
Compared to model-based controllers, model-free controllers can learn the optimal control input from the sensor measurements directly and capture more complex nonlinearity \cite{r9}. 

Encrypted control results in a heavy utilization of communication and computation resources, as encryption significantly increases the size of communicated data, and encrypted computation involves more operations than computations using plain-text data. Thus, it is important to develop a resource-effective control methods for such applications.
Event-triggered control is a method for reducing the computation and communication demands of encrypted control. In an event-triggered control system, an event-triggering unit mediates the data transmission between the system and the controller. The system output will be transmitted to the controller only when a triggering condition is satisfied. The paper \cite{r16} introduced an event-triggering mechanism for controlling a linear system using ElGamal encryption. Most existing event-triggered controls predefine triggering conditions, typically in the form of thresholding triggering policies. Model-free event-triggered control can result in smaller communication frequency between the system and controller, compared with the model-based event-triggering control. Xing \textit{et al.} \cite{modelfreetrigger1} using Q-learning to optimal the event-triggering unit for continuous-time linear systems. Yang \textit{et al.} \cite{modelfreetigger2} introduced actor critic structure in event-triggered control scheme for linear multiagent systems. Though some model-free event-triggered control have been developed, the research on model-free event-triggered control for HVAC system remains in its early stages.

\subsection{Organization}
The paper is structured as follows. Section II discusses the proposed encrypted fully model-free event-triggered HVAC control framework. This framework features a model-free cloud-based controller that computes control inputs for the HVAC system. The design of the controller is discussed in detail in this section. Section III discusses the encrypted training and implementation of model-free controller. Section IV introduces a model-free event-triggering unit that is optimally designed to minimize the communication cost between the HVAC system and the cloud. Section V presents a numerical performance evaluation of the proposed encrypted fully model-free event-triggered cloud-based HVAC control framework using the TRNSYS simulator. Lastly, Section VI concludes the paper.

\section{Encrypted fully model-free HVAC control}
\begin{figure*}[ht]
    \centering
    \includegraphics[width=5.5in]{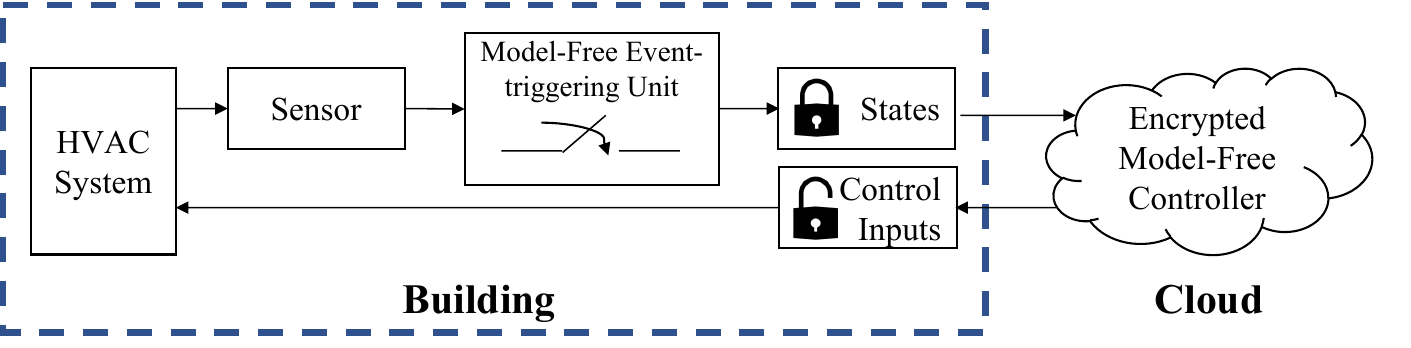}
    \caption{An overview of the encrypted fully model-free event-triggered cloud-based HVAC control framework.}
    \label{event-trigger scheme}
\end{figure*}
Protecting the occupancy information of residents in a building is a significant issue when deploying an untrusted controller, such as a cloud controller. Since the movement of residents across various areas can be inferred based on the CO$_2$ measurement in the indoor environment \cite{co2infer}. To address this problem, we propose an encrypted fully model-free approach for regulating the HVAC system, as shown in Fig. \ref{event-trigger scheme}. The training of this approach can be divided into two separately stages. In the first stage, an encrypted model-free controller is trained based on the encrypted measurements. Then the trained encrypted model-free controller is used to train the model-free event-triggering unit in the second stage. And the weights of controller are frozen in this stage. The event-triggering unit reduces communication costs by deciding whether measurements should be sent to the cloud-based controller. We chose the FHE scheme \cite{cheon2017homomorphic} to encrypt the temperature and CO$_2$ measurements, because it has the advantage of supporting both encrypted addition and multiplication which is more suitable for our case that requires complex computations. Finally, the trained encrypted model-free controller and the trained model-free event-triggering unit form a fully model-free control framework to balance communication and computation demands. 
After the training of the controller and event-triggering unit, measurements are first encrypted and, based on the event-triggering unit’s decision, sent to the controller. The controller processes these encrypted measurements and calculates the air mass flow required to regulate temperature and CO$_2$ levels. Finally, the encrypted control inputs are transmitted back to the HVAC system, where they are decrypted before being executed by the actuators.

In this section, we only discuss the design of the model-free controller. The encryption of the model-free controller and the optimal design of the model-free event-triggering unit will be discussed in the following sections.

\subsection{System Description}
Consider the following unknown building system dynamics:
\begin{equation}\label{unknown system}
    \bm{x}_{k+1}=f(\bm{x}_k,\bm{u}_k), 
\end{equation}
where $\bm{x}_k\in\mathbb{R}^{n}$ represents the $n$-dimensional vector of temperature and CO$_2$ concentration at time $k$, and $\bm{u}_k\in\mathbb{R}^{m}$ is a $m$-dimensional vector of the control input for regulating the temperature at time $k$. $\bm{x}_k$ is obtained by sensor measurements and the model-free controller generates control input $\bm{u}_k$.

\begin{align}
    \label{objective function}
     \min_{a_1,...,a_H} &\sum_{k=1}^H E\left[ (\bm{x}_k-\bm{x}_r)^\mathrm{T} \bm{Q} (\bm{x}_k-\bm{x}_r) + \bm{u}_{k}^\mathrm{T} \bm{R} \bm{u}_{k}+ \beta a_k\right]+ E\left[(\bm{x}_{H+1}-\bm{x}_r)^\mathrm{T} \bm{Q} (\bm{x}_{H+1}-\bm{x}_r)\right].
\end{align} 

For introducing the event-triggering unit into the system framework, we formulate an optimal control problem as \eqref{objective function}. $\bm{x}_r\in\mathbb{R}^{n}$ is the reference signal, and $\bm{Q}\in\mathbb{R}^{n \times n}$ and $\bm{R}\in\mathbb{R}^{m \times m}$ are two positive definite matrices. The decision of the event-triggering unit is denoted by $a_k \in [0,1]$, determines whether the system states are transmitted to the controller. Specifically, when $a_k = 1$, the system states are transmitted, model-free controller generates $\bm{u}_{k}$ at time $k$; when $a_k =0$, the system states are not transmitted, and the system executes the last control input to update the state. Finally, $\beta$ is communication penalty.

\subsection{The Design of Model-free Controller}
An optimal model-based HVAC control framework requires specific building dynamic equations to compute the control input. When dealing with complex thermal models, simplifying assumptions are often made, which can result in the omission of important details. Also, model-based controllers can be computationally demanding. For example, MPC requires solving an optimization problem to compute the control input at each step. In contrast, the proposed model-free controller in this section relies solely on sensor measurements. This approach allows for the modeling of complex building environments and can be easily adapted to various applications. 

Buildings have complex nonlinear dynamics, and each room's dynamic may influence others. Meanwhile, the control of the HVAC system can be delayed. Goal representation Heuristic Dynamic Programming (GrHDP) \cite{ni2014grdhp} introduces an internal reinforcement signal that adapts based on future external signals, rather than relying solely on the utility function, which is common in standard reinforcement learning algorithms. This approach anticipates future states and rewards, resulting in better control performance, particularly in environments with delayed rewards or complex dynamics. Q-learning, which relies on immediate rewards to update the Q-values, might not capture long-term dependencies as effectively as GrHDP, especially in environments with delayed rewards or complex dynamics. Moreover, Q-learning is not applicable to continuous action spaces because it relies on a finite Q-table where each state-action pair is explicitly represented. However, GrHDP can handle continuous action spaces by learning a deterministic policy that maps states directly to actions, thus enabling the agent to take continuous actions. Motivated by these, we choose the GrHDP method to design the model-free controller.

Based on GrHDP algorithm, the internal reinforcement signal of system in \eqref{unknown system} can be written as follows,
\begin{equation}\label{internaleq}
    G(\bm{x}_k,\bm{u}_k)=r(\bm{x}_k,\bm{u}_k)+\alpha G(\bm{x}_{k+1},\bm{u}_{k+1}),
\end{equation}
where $0 \leq \alpha \leq 1$ is a discount factor, and $r(\bm{x}_k,\bm{u}_k)$ is the external reinforcement signal which is chosen as the quadratic form
\begin{equation}
    r(\bm{x}_k,\bm{u}_k)= \bm{x}_k^\mathrm{T}\bm{D}\bm{x}_k+\bm{u}_k^\mathrm{T}\bm{M}\bm{u}_k,
\end{equation}
where $\bm{D}\in\mathbb{R}^{n \times n}$ and $\bm{M}\in\mathbb{R}^{m \times m}$ are positive definite matrices with appropriate dimensions.
The value function can be formulated as:
\begin{equation}
    W(\bm{x}_k,\bm{u}_k)=G(\bm{x}_k,\bm{u}_k)+\gamma W(\bm{x}_{k+1},\bm{u}_{k+1}),
\end{equation}
where $\gamma \in [0,1]$ is also a discount factor.
The optimal control is determined by the internal reinforcement signal $G(\bm{x}_k,\bm{u}_k)$ and value function $W(\bm{x}_k,\bm{u}_k)$. To this end, in the design of the model-free controller, three networks are needed to approximate $G(\bm{x}_k,\bm{u}_k)$, $W(\bm{x}_k,\bm{u}_k)$ and $\bm{u}_k$, respectively. 

The controller training process starts with the initial $G^0(\bm{x}_k,\bm{u}_k) = 0$, $W^0(\bm{x}_k,\bm{u}_k)=0$, and an arbitrary initial control input $\bm{u}_k$ in the training epoch 0.

Then based on Eq. \eqref{internaleq}, the internal reinforcement signal $G$ in the $p+1$-th training epoch can be updated by
\begin{equation}\label{internal}
    G^{p+1}(\bm{x}_k,\bm{u}_k^p) = r(\bm{x}_k,\bm{u}_k^p)+\alpha G^p({\bm{x}_{k+1},\bm{u}_{k+1}^p}).
\end{equation}

At the same time, value function $W$ in the $p+1$-th training epoch can be updated by
\begin{align}\label{performanceindex}
    W^{p+1}(\bm{x}_k,\bm{u}_k^p)\nonumber &= \min_{\bm{u}_k}\left\{r(\bm{x}_k,\bm{u}_k)+\alpha G^p (x_{k+1},\bm{u}_{k+1})+\gamma W^p(\bm{x}_{k+1},\bm{u}_{k+1})\right\}\nonumber\\
    &=r(\bm{x}_k,\bm{u}_k^p)+\alpha G^p (\bm{x}_{k+1},\bm{u}_{k+1}^p)+\gamma W^p(\bm{x}_{k+1},\bm{u}_{k+1}^{p}).
\end{align} 
Similarly, the control input $\bm{u}_k^{p+1}$ is updated by
\begin{equation}\label{updatecontrol}
    \bm{u}_k^{p+1} = \arg \min_{\bm{u}_k}W^{p+1}(\bm{x}_k,\bm{u}_k^{p}).
\end{equation}
Repeat executing Eqs. \eqref{internal}-\eqref{updatecontrol} until convergence.

To implement the above GrDHP algorithm, we use three neural networks to model the internal reinforcement signal $G$ (Critic network 1), the value function $W$ (Critic network 2), and the controller (Actor).
\begin{figure*}[ht]
    \centering
    \includegraphics[width=5.5in]{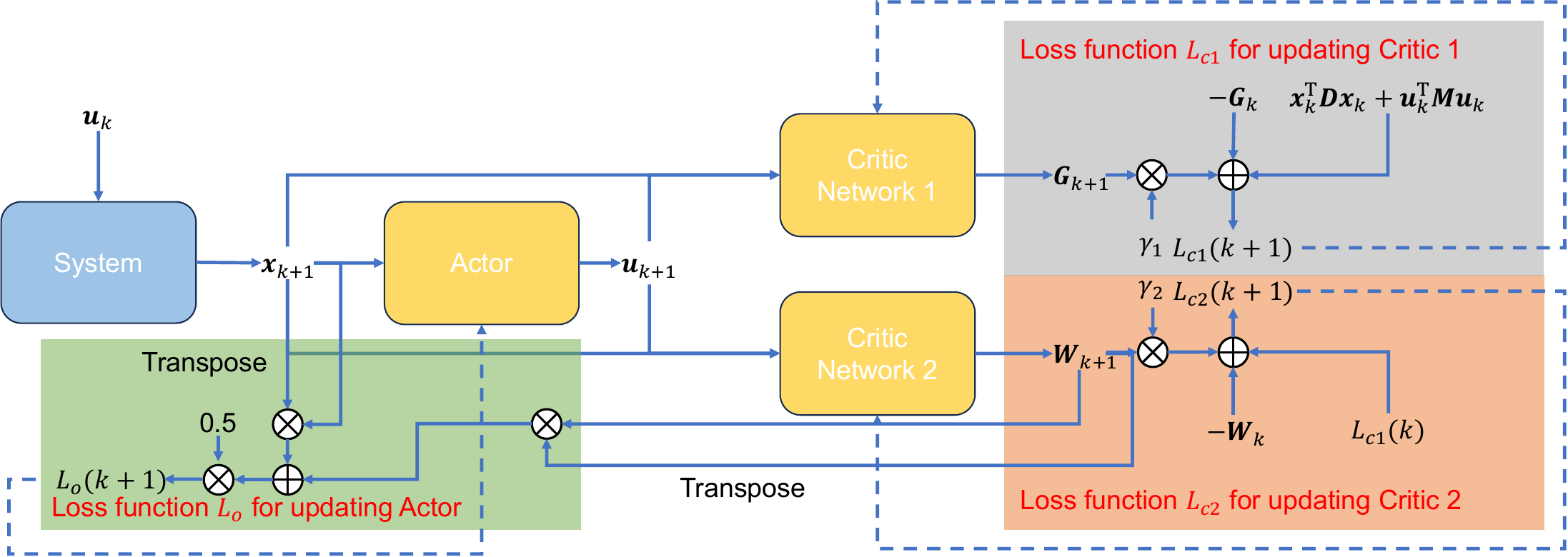}
    \caption{The forward and backward process for training the model-free controller.}
    \label{controller network}
\end{figure*}

\subsubsection{Actor}
The controller or actor is composed of two linear layers as follows,
\begin{align}
    \rho_k = \sigma_1 \mathfrak{W}_{a1} \bm{x}_k, \nonumber \\
    \bm{u}_k = \textbf{\textit{u}}_{max}\cdot\sigma_2 \mathfrak{W}_{a2}\rho_k, \nonumber 
\end{align}
where $\mathfrak{W}_{a1} \in \mathbb{R}^{d \times n}$ and $\mathfrak{W}_{a2} \in \mathbb{R}^{m \times d}$ are the weight matrices, $d$ is the number of hidden units, $\cdot$ denotes the element-wise multiplication, $\sigma_1$ is the Tanh activation function, $\sigma_2$ is the Sigmoid activation function, $\textbf{\textit{u}}_{max}$ is the upper bound of $\textbf{\textit{u}}$, $\rho_k$ is the output of the first layer. Note that the purpose of the Sigmoid function is to constrain the output into $[0,1]$, because in our case, the control input is non-negative. 

According to Eq. \eqref{updatecontrol}, the loss function of the actor is defined as
\begin{equation}
    L_o =  \min \frac{1}{2} W^T(\bm{x}_k,\bm{u}_k)W(\bm{x}_k,\bm{u}_k). \nonumber
\end{equation}

The weights of the actor are updated according to back propagation algorithm and the chain rule as follows,
\begin{align}
    \mathfrak{W}^{'}_{a2} &= \mathfrak{W}_{a2}-\zeta_a(\frac{\partial L_o}{\partial W(\bm{x}_k,\bm{u}_k)} \frac{\partial W(\bm{x}_k,\bm{u}_k)}{\partial \bm{u}_k} \frac{\partial \bm{u}_k}{\partial \mathfrak{W}_{a2}}), \nonumber \\
    \mathfrak{W}^{'}_{a1} &= \mathfrak{W}_{a1}-\zeta_a(\frac{\partial L_o}{\partial W(\bm{x}_k,\bm{u}_k)} \frac{\partial W(\bm{x}_k,\bm{u}_k)}{\partial \bm{u}_k} \frac{\partial \bm{u}_k}{\partial \rho_k} \frac{\partial \rho_k}{\partial \mathfrak{W}_{a1}}), \nonumber
\end{align}
where $\zeta_a>0$ is the learning rate for the actor, and $\mathfrak{W}^{'}$ denotes the updated weight matrix.
\subsubsection{Critics}
Two critics use a common two-layer fully connected neural network architecture as follows,
\begin{align}
    h_k &= \sigma_2 \mathfrak{W}_{c1} {\rm{cat}}(\bm{x}_k,\bm{u}_k), \nonumber \\
    S_k &= \mathfrak{W}_{c2}h_k, \nonumber 
\end{align}
where $\mathfrak{W}_{c1}\in \mathbb{R}^{d \times (m+n)}$ and $\mathfrak{W}_{c2}\in \mathbb{R}^{m \times d}$ are two weight matrices, $\rm{cat}$ is the concatenation function, and $h_k$ is the output of the first layer.

Critic network 1 outputs $S_k$ as the internal reinforcement signal $G(\bm{x}_k,\bm{u}_k)$ with the input ${\rm{cat}}(\bm{x}_k,\bm{u}_k)$. The loss function of critic network 1 is defined according to Eq. \eqref{internal} as follows,
\begin{equation}
    L_{c1}= \min \frac{1}{2}(r(\bm{x}_k,\bm{u}_k)+\alpha G (\bm{x}_{k+1},\bm{u}_{k+1})-G^{'}(\bm{x}_k,\bm{u}_k))^2. \nonumber
\end{equation}

Critic network 2 is used to approximate the value function $W$. Its loss function is defined according to Eq. \eqref{performanceindex} as follows,
\begin{equation}
    L_{c2}=\min \frac{1}{2} (G^{'}(\bm{x}_k,\bm{u}_k)+\gamma W (\bm{x}_{k+1},\bm{u}_{k+1})-W^{'}(\bm{x}_k,\bm{u}_k))^2. \nonumber
\end{equation}

According to back propagation algorithm and the chain rule, the update of weights of two critics can use a common formula to represent:
\begin{align}
    \mathfrak{W}_{c2}^{'} &= \mathfrak{W}_{c2}-\zeta(\frac{\partial L_c}{\partial S(\bm{x}_k,\bm{u}_k)} \frac{\partial S(\bm{x}_k,\bm{u}_k)}{\partial \mathfrak{W}_{c2}}), \nonumber \\
    \mathfrak{W}_{c1}^{'} &= \mathfrak{W}_{c1}-\zeta(\frac{\partial L_c}{\partial S(\bm{x}_k,\bm{u}_k)} \frac{\partial S(\bm{x}_k,\bm{u}_k)}{\partial h_k}\frac{\partial h_k}{\partial \mathfrak{W}_{c1}}), \nonumber
\end{align}
where $L_c$ denotes the loss function of each critic network. Fig. \ref{controller network} shows the forward and backward process for training the proposed model-free controller.

\begin{rmk}
In the learning process of the controller, detailed information about the system functions is not required. This is important, as obtaining exact information about system functions is often challenging in many practical scenarios.
\end{rmk}

\section{Secure Training and Implementation of Model-Free HVAC Controller}
For a model-free cloud-based controller, there is a potential risk of privacy loss when the controller is not fully trusted. To mitigate this risk, encryption schemes can be used to protect sensitive information during the training and implementation stages. In encrypted training of a model-free controller, the states or measurements of the system are encrypted on the client side and sent to the model-free controller which resides on the cloud. The model-free controller then computes the desired control input using the encrypted data. The encrypted control input is sent back to the client for decryption and execution. We use GrHDP to train the encrypted model-free controller. It is important to note that the weights of each network in the controller are not encrypted.

Under encrypted measurements, the calculations of control input differ significantly from calculations involving only plaintext. This difference arises because encryption transforms data into a format that is not directly interpretable without decryption. As a result, standard arithmetic operations cannot be directly applied to encrypted measurements. Instead, operations on ciphertext must be conducted using specific cryptographic techniques, such as HE. In the secure training, three neural networks are involved: the actor (controller) and two critics, which consist of linear layers and activation functions. Therefore, multiple operations between encrypted vectors/matrices and plain matrices are involved in the encrypted training process. To handle these operations, we use the CKKS encryption scheme and the TenSEAL library to encrypt the states or measurements of the system. More details about CKKS can be found in Appendix \ref{App: CKKS}.
\begin{figure}[h]
    \centering
    \includegraphics[width=3.4in]{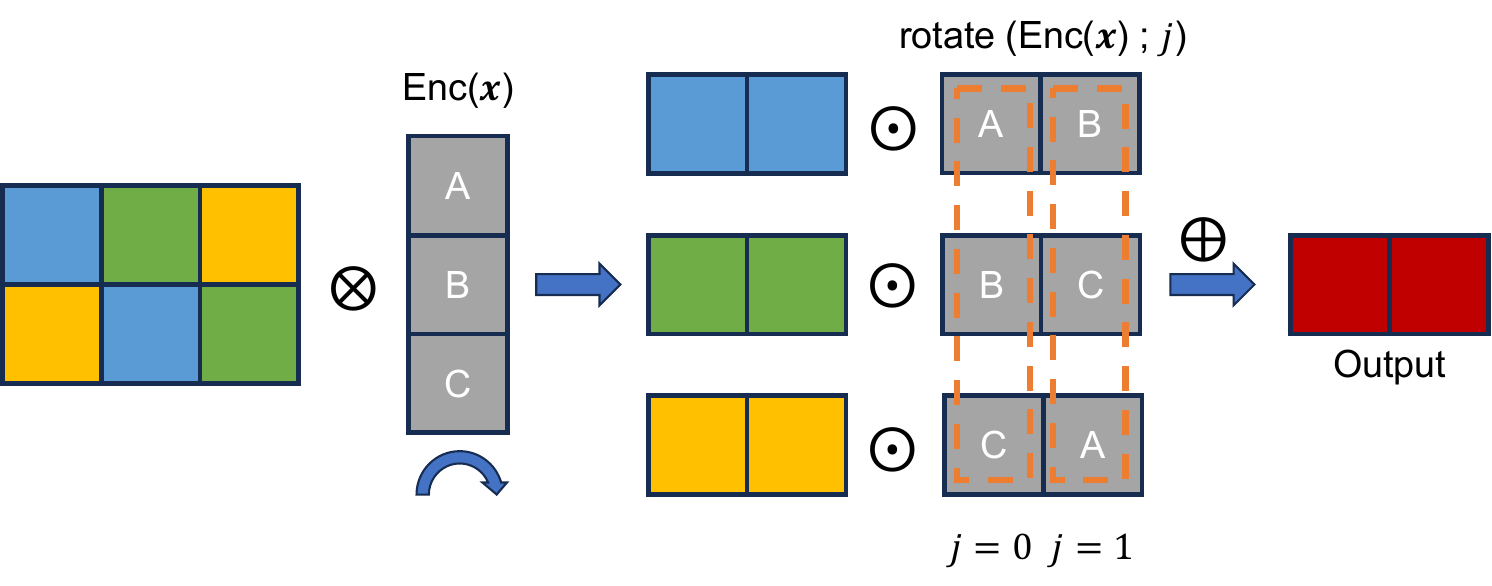}
    \caption{Encrypted matrix-vector multiplication using the diagonal method.}
    \label{fig:packing}
\end{figure}
\subsection{Encrypted Implementation of Linear Layers} 
The linear layer consists of matrix-vector/matrix multiplication and matrix-vector/matrix addition. Matrix-vector/matrix multiplication is based on packing and rotation techniques for the purpose of acceleration. The encryption for a linear layer can be represented as a sequence of operations, including additions, multiplications, and rotations over packed ciphertexts, while the weight matrices of the linear layer remain unencrypted. Packing refers to the technique of combining multiple data elements into a single ciphertext, allowing parallel computation on encrypted data. Rotation involves shifting the elements within the packed ciphertext, enabling the interaction between different data elements during encrypted computations, which is essential for matrix operations like convolutions or multiplications in neural networks. Tenseal uses Gazelle diagonal method \cite{juvekar2018gazelle} to implement matrix-vector multiplication. Fig. \ref{fig:packing} shows matrix-vector multiplication between an encrypted gray vector and an unencrypted weight matrix using the diagonal method.

Given a weight matrix $\mathfrak{W}$ and the system state $\bm{x}=[A,B,C]^\mathrm{T}$, encrypted matrix-vector multiplication is done as shown in Fig. \ref{fig:packing}. In this case, input $\bm{x}$ is ciphertext which is denoted as $\enc{\bm{x}}$. We then perform iteration which multiplies each of packed diagonal component of the weight $\mathfrak{W}$ with rotated packed input vector. Packed $\enc{\bm{x}}$ is rotated by one at each iteration step. Note that $rotate(\enc{\bm{x}};j)$ denotes $\enc{\bm{x}}$ rotated to upward direction by $j$ slots. Once the iterations are complete, the sum of results produces a linear transformation. As the rotation is solely performed on the input vector, the computation can be reduced. Consequently, the overall computational expense consists of $n$ multiplications and $n$ rotations. The procedure is outlined in algorithm \ref{packing_algorithm}.

\begin{algorithm}[ht]
\caption{Encrypted matrix-vector multiplication}  \label{packing_algorithm}
\KwIn{$\mathfrak{W}= \left\{\mathfrak{w}_{i,j},0\leq i\leq d-1, 0\leq j\leq n-1\right\}$, $\enc{\bm{x}}$}
\KwOut{$\mathfrak{Y}$}
Set empty matrices $\bm{q},\bm{p} \in \mathbb{R}^{n \times d} $ \;
\For{$j=0$ \KwTo $n-1$}{
  $\bm{q}^{col}_j = \textit{rotate}(\enc{\bm{x}}; j) $ \;
  $\bm{p}^{row}_{j} = [\mathfrak{w}_{0,j}, \mathfrak{w}_{1,j+1},\ldots, \mathfrak{w}_{d,(j+d-1)\mod n}]$ \; 
}
$\mathfrak{Y} = \sum_{j=0}^{j=n-1} \bm{q}^{row}_{j} \cdot \bm{p}^{row}_{j}$.
\end{algorithm}

\subsection{Encrypted Implementation of Activation}
HE supports only polynomial arithmetic operations. The Sigmoid and Tanh functions pose a challenge for evaluation because they cannot be represented as polynomials. The Sigmoid and Tanh function for an input $\bm{y}$ are defined as,
\begin{align}
    Sig(\bm{y}) &= \frac{1}{1+e^{-\bm{y}}}, \nonumber\\
    Tanh(\bm{y}) &= \frac{e^{\bm{y}}-e^{-\bm{y}}}{e^{\bm{y}}+e^{-\bm{y}}}.\nonumber
\end{align}
To address this problem, we use the least squares approach to find a polynomial approximation of the Sigmoid and Tanh function. The resulting approximate polynomial $A_{Sig}(\bm{y})$ and $A_{Tanh}(\bm{y})$ of degree $5$ for Sigmoid and Tanh functions are given by:
\begin{align}
    A_{Sig}(\bm{y}) &= \theta_0 + \theta_1\bm{y} + \theta_3\bm{y}^3 + \theta_5\bm{y}^5, \nonumber \\
    A_{Tanh}(\bm{y}) &= \delta_0 + \delta_2\bm{y}^2 + \delta_4\bm{y}^4 + \delta_5\bm{y}^5, \nonumber
\end{align}
where coeffcient vectors are $(\theta_0,\theta_1,\theta_3,\theta_5) \approx (0.5,-1.53,2.35,-1.35)$ and $(\delta_0,\delta_2,\delta_4,\delta_5) \approx (0.024,-0.212,
  0.958, -0.002)$.

Based on these operations, it is clear that each network needs exactly three encrypted multiplications: one for the first linear layer, one for the activation, and one for the final linear layer.

\section{The Design of Model-Free Event-triggering Unit}
To reduce the high communication and computation costs associated with FHE, we introduce a model-free event-triggering unit that determines when the system should transmit encrypted measurements to the cloud, as illustrated in Fig. \ref{event-trigger scheme}. For convenience, we separate the training of the model-free event-triggering unit from the training of the model-free controller. The event-triggering unit is trained using the trained controller, and the controller weights are fixed in the training of the event-triggering unit.

In most event-triggered optimal control problems, event-triggering conditions are designed based on system dynamics. In this section, we propose a model-free design for the event-triggering unit, where the triggering condition is estimated by a neural network based on state measurements. This method effectively reduces the frequency of triggering instances and the computational demand of the encrypted model-free controller.
 
At each time $k$, the event-triggering unit makes a decision based on the available information $\mathfrak{F}_k$. Initially, we set $a_0(\mathfrak{F}_0)=1$ at time $k=0$. To ensure the event-triggering unit does not perpetually avoid communication, we introduce a constraint on the communication frequency. Let $N_k$ represent the number of samples collected since the last communication between the system and the controller, that is 
\begin{equation}
\label{L_t}
N_k = k - k^\star,\nonumber
\end{equation}
where $k^\star$ is the last time instance when the event-triggering unit communicated the states to the controller. If $N_k>T_s$, where $T_s$ is a threshold, the event-triggering unit allows the system to communicate with the controller. This constraint ensures system states are transmitted at least once in every $T_s$ time step, maintaining effective control.

The optimal triggering policy can be obtained by solving the optimal control problem in \eqref{objective function}. 
The event-triggering unit incurs a penalty of $\beta$ if $a_k=1$, causing the HVAC system to execute $\bm{u}_{k}$ at time $k$. If there is no communication ($a_k=0$), the system applies $\bm{u}_{k^\star}$ to update $\bm{x}_{k+1}$, where $\bm{u}_{k^\star}$ the control input which was computed at time $k^\star$ using $\bm{x}_{k^\star}$. This results in the optimization problem in \eqref{objective function} being non-Markovian. Since existing Markov Decision Process (MDP) algorithms only determine optimal policies for Markovian problems, we introduce new states to transform \eqref{objective function} into an MDP.

Define $\bm{z}_k$ as a new state which is updated according to \begin{equation}
\label{state_y}
\bm{z}_{k+1} = (1 - a_k)\bm{z}_k + a_k\bm{x}_k,\nonumber
\end{equation}
where $z_t$ is initialized as $\bm{z}_1 = \bm{x}_1$. Note that $\bm{z}_k$ stores the last communicated state with the cloud. If $a_k=1$, we have $\bm{z}_k = \bm{x}_k$ which stores the system state at time $k$. When $a_k=0$, we have $\bm{z}_k = \bm{x}_{k^\star}$ which stores the system state at the time of the last communication. Next, we introduce the new state $z_k$ to $\mathfrak{F}_k$, \emph{i.e.}, the available information at the event-triggering unit, which can be updated as follows:
\begin{equation}
\begin{aligned}
\begin{bmatrix} \bm{x}_{k+1} \\
\bm{z}_{k+1} \\
N_{k+1}\end{bmatrix}= 
\begin{bmatrix} f(\bm{x}_k,\bm{u}_k) \\
(1 - a_k)\bm{z}_k + a_k\bm{x}_k \\
(1 - a_k)(N_k+1)\end{bmatrix}.
\end{aligned}
\label{It}
\end{equation}
Note that $\mathfrak{F}_{k+1}$ is only related to $\mathfrak{F}_k$, the dynamics in \eqref{It} is Markovian.

Then the cost function (total cost) $C_k$ and value function $V_k$ for the event-triggering unit design problem can be defined as
\begin{equation}
\label{cost}
    C_k (\mathfrak{F}_k,a_k ) = (\bm{x}_k-\bm{x}_r)^\mathrm{T} \bm{Q} (\bm{x}_k-\bm{x}_r) + (1 - a_k)\bm{u}_{k^\star}^\mathrm{T} \bm{R} \bm{u}_{k^\star} + a_k\bm{u}_{k}^\mathrm{T} \bm{R} \bm{u}_{k} + \beta a_k,
\end{equation}
\begin{equation}
\label{valuec}
    V_k (\mathfrak{F}_k) = \min_{a_k \in A_k}(C_t (\mathfrak{F}_k,a_k ) + E \left\{ V_{k+1} (\mathfrak{F}_{k+1})|\mathfrak{F}_k,a_k \right\} ),
\end{equation}
where $A_k=\left\{0,1\right\}$ is the action set at time $k$. The objective function in an MDP is to find a policy $\pi$ to minimize value function $V_k (\mathfrak{F}_k)$, so (\ref{valuec}) can be written as follows:
\begin{equation}
\label{minvalue}
V_k^\pi (\mathfrak{F}_k) = C_k (\mathfrak{F}_k,a_k^\pi ) + E \left\{ V_{k+1}^\pi (\mathfrak{F}_{k+1})|\mathfrak{F}_k,a_k^\pi \right\}. \nonumber
\end{equation}

We initialize the terminal $V_{H+1} (\mathfrak{F}_{H+1})$ as
\begin{equation}
\label{terminal}
V_{H+1} (\mathfrak{F}_{H+1}) = (\bm{x}_{H+1}-\bm{x}_r)^\mathrm{T} \bm{Q} (\bm{x}_{H+1}-\bm{x}_r). \nonumber
\end{equation}

Using the backward recursion equation (\ref{valuec}), we can obtain the solution to the finite horizon optimal control problem as follows:
\begin{equation}
\label{value}
\begin{aligned}
V_H (\mathfrak{F}_H) &= \min_{a_H \in A_H}(C_H (\mathfrak{F}_H,a_H ) + E \left\{ V_{H+1} (\mathfrak{F}_{H+1})|\mathfrak{F}_H,a_H \right\} ),\\
&\vdots \\
V_1 (\mathfrak{F}_1) &= \min_{a_1 \in A_1}(C_1 (\mathfrak{F}_1,a_1 ) + E \left\{ V_{2} (\mathfrak{F}_{2})|\mathfrak{F}_1,a_1 \right\} ). \nonumber
\end{aligned}
\end{equation}

The equation \eqref{valuec} offers a recursive optimality equation. It can be utilized in different reinforcement learning methods where the optimal event-triggering policy can be obtained. In standard MDP formulations, the optimal value function depends solely on the current state. However, in our case, the necessary information set is defined by $\mathfrak{F}_k=[\bm{x}_{k},z_{k},N_{k}]$, the optimal value function at time $k$ is a function of it, which is different from standard MDP. 

\section{Numerical Results}
\subsection{Simulation Set-up}
\begin{figure}[h]
    \centering
    \includegraphics[width=2.6in]{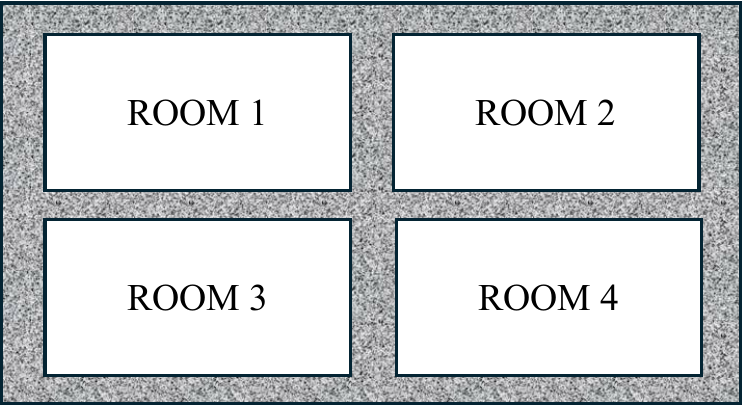}
    \caption{The outlay of rooms in the building. Note that the shadow parts are the walls of the building.}
    \label{room}
\end{figure}

In this section, we evaluate the performance of the proposed encrypted fully model-free event-triggered control framework utilizing the TRNSYS building simulator. For this purpose, we modeled a single-story house within TRNSYS, as illustrated in Fig. \ref{room}. The building spans a total floor area of $3000$ square feet and includes four rooms. The HVAC system within the building regulates both CO$_2$ concentration and temperature. Each room has a maximum of eight individuals, varying occupancy according to weekday and weekend schedules. The simulation used weather data from Fresno City for July $2016$, with a reference temperature set at $23.5$ degrees Celsius and a reference CO$_2$ level of 800 ppm.

In the training of the model-free controller, the system states are encrypted. The training process follows a sequential order: first critic 1, then critic 2, and finally the actor. Initially, the weights $\mathfrak{W}_{(g,c1)}$ and $\mathfrak{W}_{(g,c2)}$ of critic 1 are trained and subsequently fixed. The training of the actor needs the interaction of critic 2 (see Eq. \ref{updatecontrol}). Hence, in the training of the actor, the weights of critic 2 are frozen. The model-free event-triggering unit is trained after the training of model-free controller. Specifically, the pretrained actor (controller) is used to train the event-triggering unit and the weights of actor are frozen.

We developed and evaluated the proposed fully model-free control framework using PyTorch in Python, with the CKKS encryption and decryption based on Tenseal. Based on the official benchmark of TenSEAL, we configured the CKKS parameters with a polynomial degree of $N=8192$, coefficient modulus sizes $q = [40, 26, 26, 26, 40]$, and a scaling factor set to $2^{26}$. Communication between TRNSYS and Python was facilitated using TRNSYS type 3157. All simulations were conducted in Python and TRNSYS on an Intel Core i7-10700 CPU 2.9 GHz computer.

\subsection{Benchmark}
We compare the performance of the proposed encrypted fully model-free event-triggered control framework with an encrypted model-based event-triggered control method where the MPC method is used to regulate the indoor CO$_2$ and temperature.
Under the model-based approach, two fully connected layers are used to optimize the event-triggering policy by using the REINFORCE algorithm \cite{williams1992simple}. To ensure fairness, the optimal event-triggering policy of the proposed encrypted fully model-free event-triggered control framework is also obtained using the REINFORCE algorithm. The training settings and hyperparameters of the encrypted model-based event-triggered control framework \cite{feng2023privacypreserving} are kept the default according to \cite{feng2023privacypreserving}. The measurements for the encrypted model-based event-triggered control framework \cite{feng2023privacypreserving} are also encrypted and decrypted by the CKKS scheme.

In the following subsections, the optimal fully model-free policy represents the policy obtained by the proposed encrypted fully model-free event-triggered control framework, whereas, the model-based policy denotes the policy from the encrypted model-based event-triggered control framework \cite{feng2023privacypreserving}.

\subsection{Assessment Metrics}
 An ideal indoor living condition requires room temperature and CO$_2$ concentration to stay in the comfort range. Therefore, control performance of the HVAC system is evaluated in terms of both percentage of violations and maximum deviation which are defined as follows. The proportion of time during which the temperature in any room exceeds the comfort range of $22-25$ Celsius is referred to as the temperature violation percentage. For CO$_2$ concentration, this percentage reflects the duration when the concentration surpasses $800$ ppm in any room. The greatest deviation of temperature from the specified comfort range, considering all rooms and all time intervals, is termed the maximum temperature deviation. Similarly, the maximum CO$_2$ deviation represents the highest recorded level exceeding the 800 ppm threshold across all rooms and times.

 Additionally, we assess the impact of the proposed fully model-free framework and the model-based framework \cite{feng2023privacypreserving} on communication frequency between the system and the cloud using the communication rate. The communication rate is defined as the percentage of time instances in which the event-triggering unit interacts with the cloud-based controller.
\begin{figure*}[ht]
\centering
\subfigure[Temperature violations obtained by optimal fully model-free and optimal model-based policies using different communication rate.]{\label{violation_temp}\includegraphics[width=2.8in]{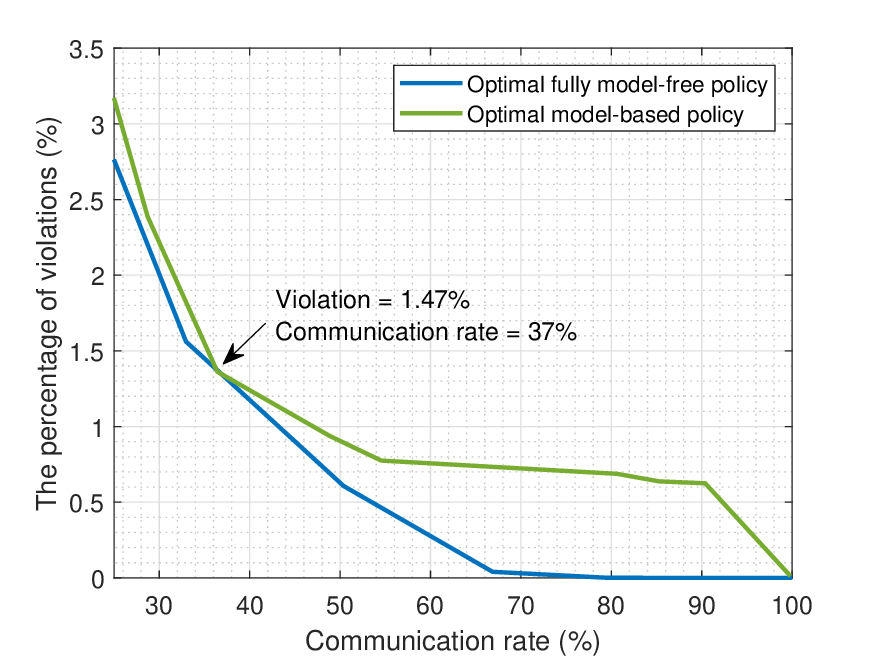}}
\quad
\subfigure[Maximum temperature deviations obtained by optimal fully model-free and optimal model-based policies using different communication rate.]{\label{deviation_temp}
\includegraphics[width=2.8in]{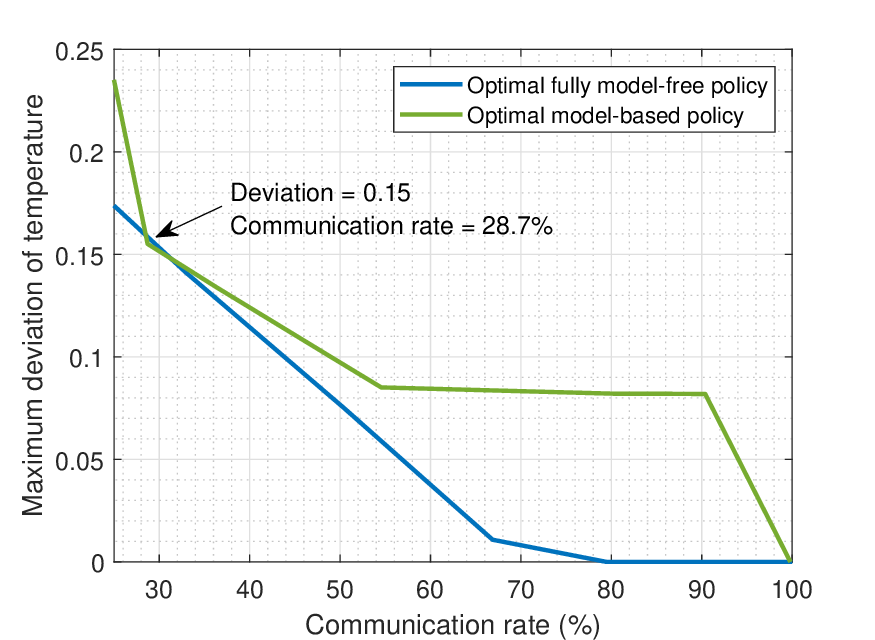}
}
\quad
\subfigure[CO$_2$ violations obtained by optimal fully model-free and optimal model-based policies using different communication rate]{\label{violation_co2}
\includegraphics[width=2.8in]{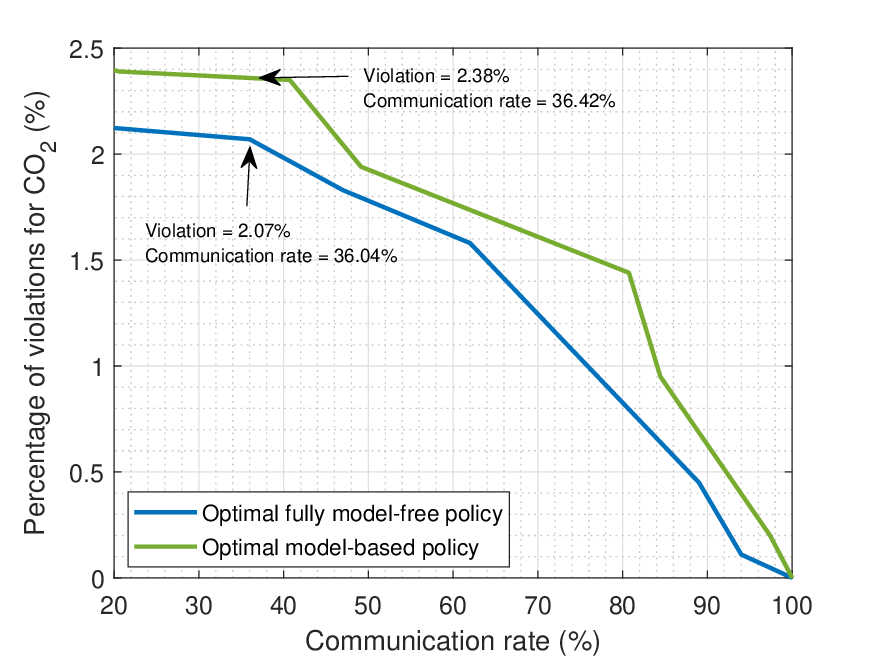}
}
\quad
\subfigure[Maximum CO$_2$ deviations obtained by optimal fully model-free and optimal model-based policies using different communication rate]{\label{deviation_co2}
\includegraphics[width=2.8in]{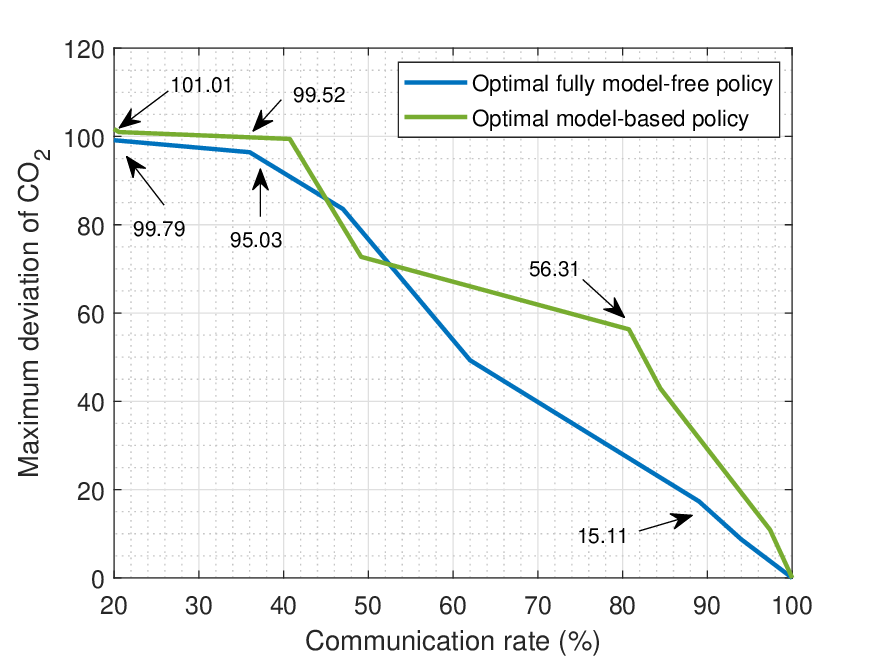}
}
\caption{The control performance of temperature and CO$_2$ obtained by the proposed optimal fully model-free policy and the optimal model-based policy.}
\end{figure*}

\subsection{Communication Rate and Control Performance Trade-off}
In this subsection, we explore the balance between control performance and communication rate. Fig. \ref{violation_temp} shows the percentage of temperature violations for both optimal model-based and model-free policies, depending on the communication rate. Various communication rates were achieved by adjusting $\beta$ (the communication penalty factor of the proposed method, see Eq. \eqref{objective function}). For each communication rate, we evaluated violations based on the HVAC system's regulation of CO$_2$ levels and temperature over a 30-day period.

As depicted in Fig. \ref{violation_temp}, the temperature violation percentage rises with the decrease in communication rate under both optimal policies. This occurs because the controller receives state updates less frequently as the communication rate drops. Consequently, the HVAC system relies on control inputs derived from outdated data. Therefore, these control inputs become less efficient in managing the temperature as the communication rate diminishes. When the communication rate is larger than $37$\%,  the percentage of violations is significantly lower under the proposed optimal fully model-free policy than the optimal model-based policy. This advantage is particularly evident when the communication rate is greater than $65$\%, and the percentage of violations under the optimal fully model-free policy remains at $0$.   

Fig. \ref{deviation_temp} shows the maximum deviation of temperature under the optimal triggering policies as a function of the communication rate. According to this figure, the maximum deviations of both methods are very small, even when the communication rate is around 25\%, the maximum deviation is less than 0.25. The maximum deviation of the optimal fully model-free policy obtained by the proposed fully model-free method is roughly the same as that of the optimal model-based policy, but the difference between the two becomes larger when the communication rate is high. For instance, the maximum deviation under the proposed optimal fully model-free policy is $0.1$ less than that under optimal model-based policy when the communication rate is around $70$\%.

The figures \ref{violation_co2} and \ref{deviation_co2} display the percentage of CO$_2$ violations and the maximum deviations, respectively. These figures indicate that as the communication rate decreases, both the percentage of CO$_2$ violations and the maximum deviation for CO$_2$ rise. Nonetheless, the optimal fully model-free policy obtained by the proposed fully model-free method leads to a lower CO$_2$ violation compared to the optimal model-based policy.
\begin{figure}[h]
    \centering
    \includegraphics[width=2.8in]{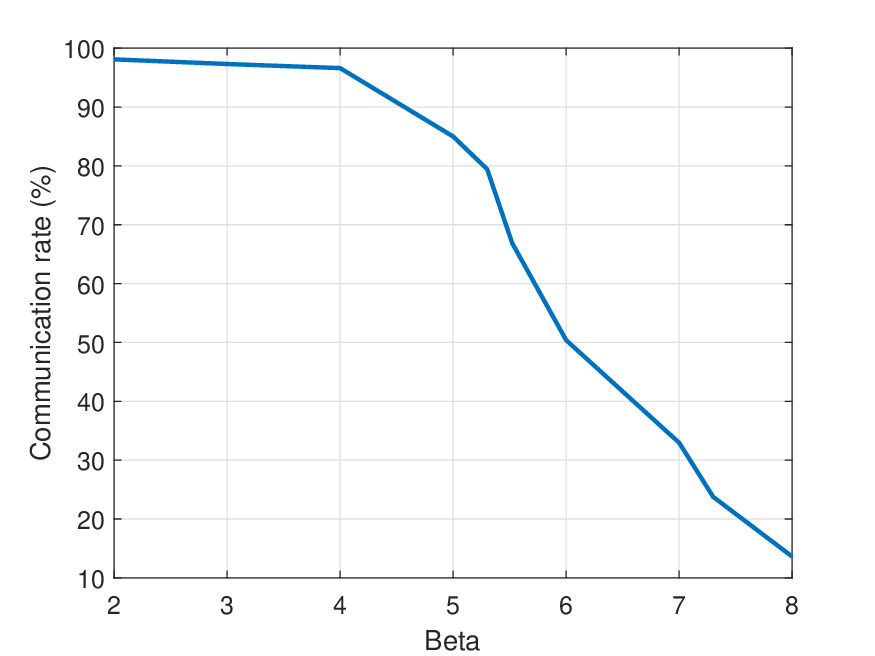}
    \caption{The communication rate of the proposed optimal fully model-free policy as a function of $\beta$.}
    \label{communication rate}
\end{figure}
Finally, Fig. \ref{communication rate} shows the communication rate of the proposed optimal fully model-free policy as a function of the communication penalty ($\beta$). As shown in the figure, the event-triggering unit reduces the frequency of communication with the cloud as the communication penalty increases, leading to a higher percentage of violations and greater maximum deviations. From Figs. \ref{violation_temp}-\ref{deviation_co2}, we can see that when the communication rate is lower than 36.04\%, the temperature violation shows a rapid upward trend, and the percentage of violations and maximum deviation for CO$_2$ also show an inflection point. We believe that the value of $\beta$ here can achieve a balance between communication and control performances.
\begin{figure}[h]
    \centering
    \includegraphics[width=2.8in]{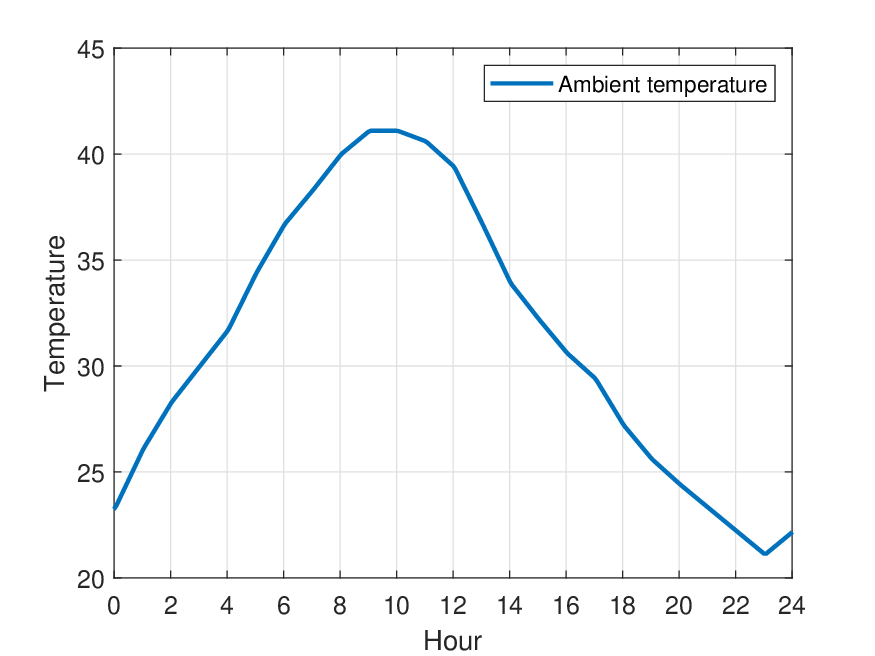}
    \caption{One day environment temperature.}
    \label{ambient}
\end{figure}
\subsection{Performance Analysis Under a Certain Communication Rate}
In this subsection, we examine the effectiveness of the HVAC system in controlling temperature and CO$_2$ levels over a one-day period using both optimal policies with a communication rate of $36\%$. Fig. \ref{ambient} illustrates the ambient temperature throughout the day, which was utilized in our simulations. This specific day is especially appropriate for analyzing the performance of event-triggering policies due to the significant temperature fluctuations causing severe disturbances.   

\begin{figure}[h]
    \centering
    \subfigure[]{\label{temthres}
    \includegraphics[width=2.8in]{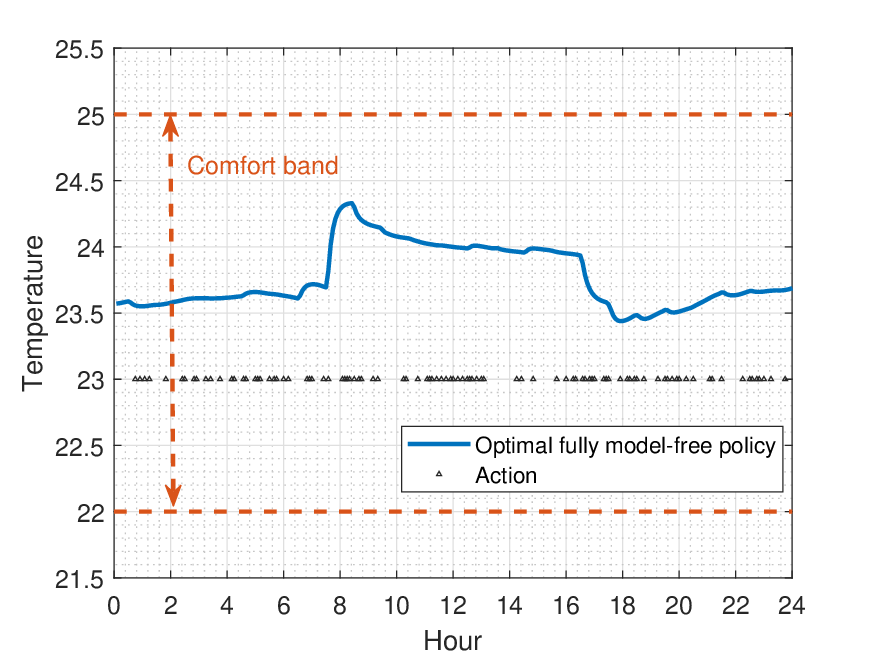}}
    \quad
    \subfigure[]{\label{temrein}
    \includegraphics[width=2.8in]{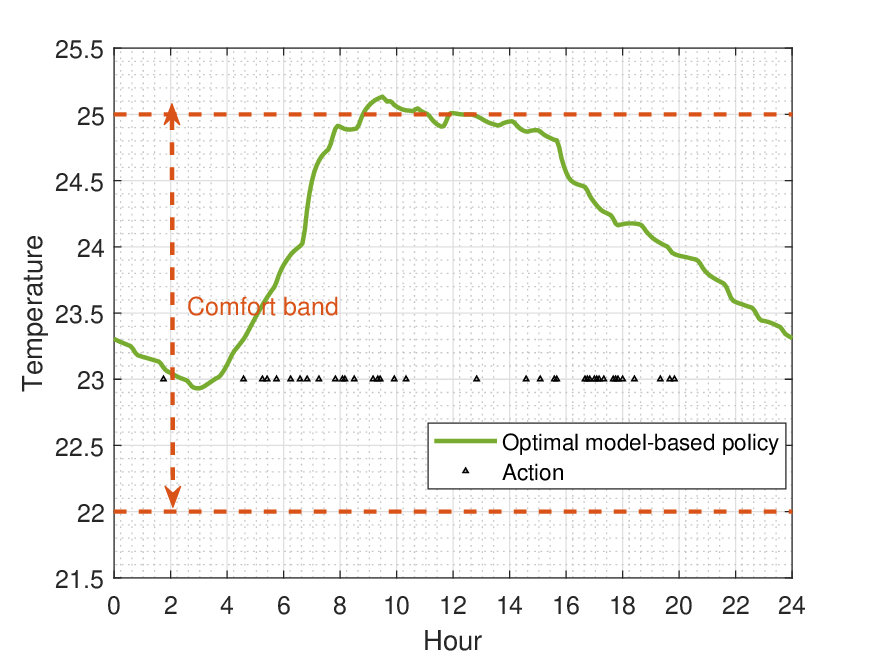}}
    \caption{The room $1$ temperature of the building under optimal fully model-free and optimal model-based policies.}
    \label{daytemall}
\end{figure}

Fig. \ref{daytemall} illustrates the temperature in room $1$ of the building under both optimal event-triggering and control policies as the outdoor temperature varies as depicted in Fig. \ref{ambient}.

From these two figures, we can see that both policies can keep the temperature within a comfortable range. When the ambient temperature reaches the highest, the optimal policy 2 can keep the temperature within $25$ Celsius, but it is close to the critical value. In contrast, the optimal policy 1 can keep the temperature around $23.5$ Celsius. Therefore, the proposed model-free method is more effective than the model-based method \cite{feng2023privacypreserving}.

The triggering instance of the event-triggering unit under both optimal policies are also shown in Figs. \ref{temrein} and \ref{temthres}, respectively. 
Under the optimal model-based policy, the event-triggering unit triggers more often when the ambient temperature changes rapidly. This causes the temperature to exceed the comfort band sometimes. However, under the optimal fully model-free policy, the distribution of the trigger time is more uniform than when using the optimal model-based policy. Therefore, under optimal fully model-free policy, the temperature fluctuation throughout the day is smaller.
\begin{figure}[h]
    \centering
    \includegraphics[width=2.8in]{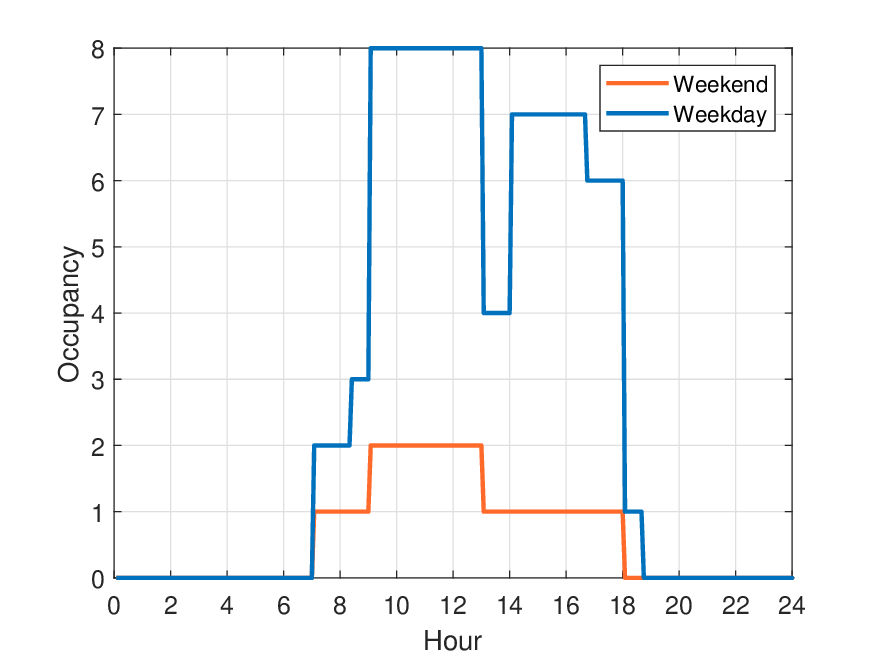}
    \caption{The weekend and weekday schedules of the occupancy in room 1.}
    \label{roompeo}
\end{figure}

\begin{figure}
    \centering
    \includegraphics[width=2.8in]{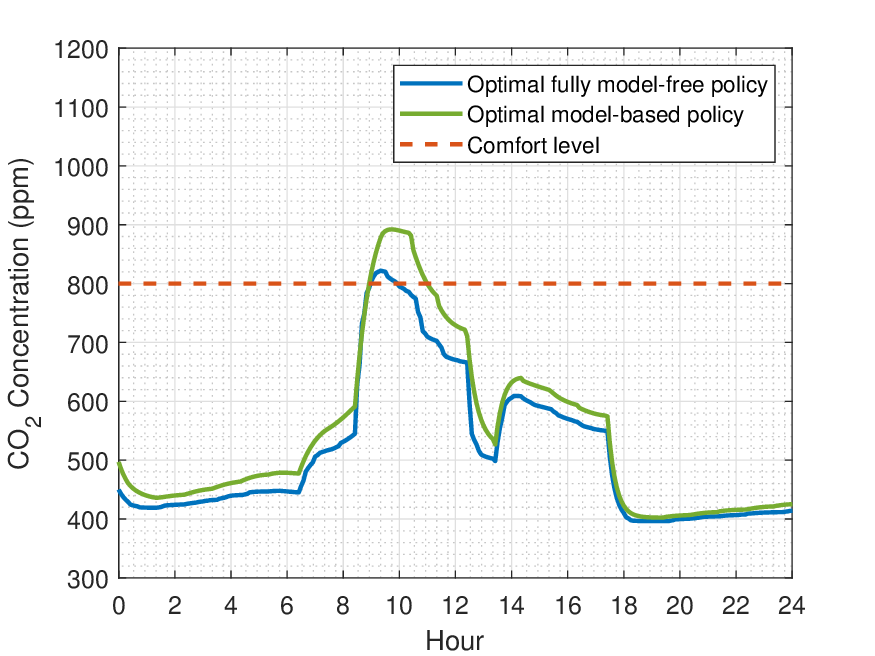}
    \caption{The CO$_2$ concentration level of room 1 under both optimal triggering policies.}
    \label{CO2graph}
\end{figure}

We subsequently examine the indoor CO$_2$ dynamics in room $1$ of the building under different event-triggering and control policies. The change of number of people in each room is shown in Fig. \ref{roompeo}. Weekdays and weekends have different schedules. Note that not every weekend in a month will have people in the building, that is, the schedule will not be executed on some weekends.
Fig. \ref{CO2graph} illustrates the CO$_2$ behavior in room $1$ under both optimal policies. As depicted in this figure, the indoor CO$_2$ concentration rises with an increase in occupancy. Under both policies, when the number of people in the room changes rapidly, CO$_2$ concentration exceeds the comfort level. However, compared with the optimal model-based policy, the maximum deviation of the optimal fully model-free policy is smaller. 
This observation confirms that the proposed model-free method is more efficient in regulating the indoor CO$_2$ concentration.

\subsection{Communication and Computation Performance}
\begin{figure}[h]
   \centering
   \includegraphics[width=3in]{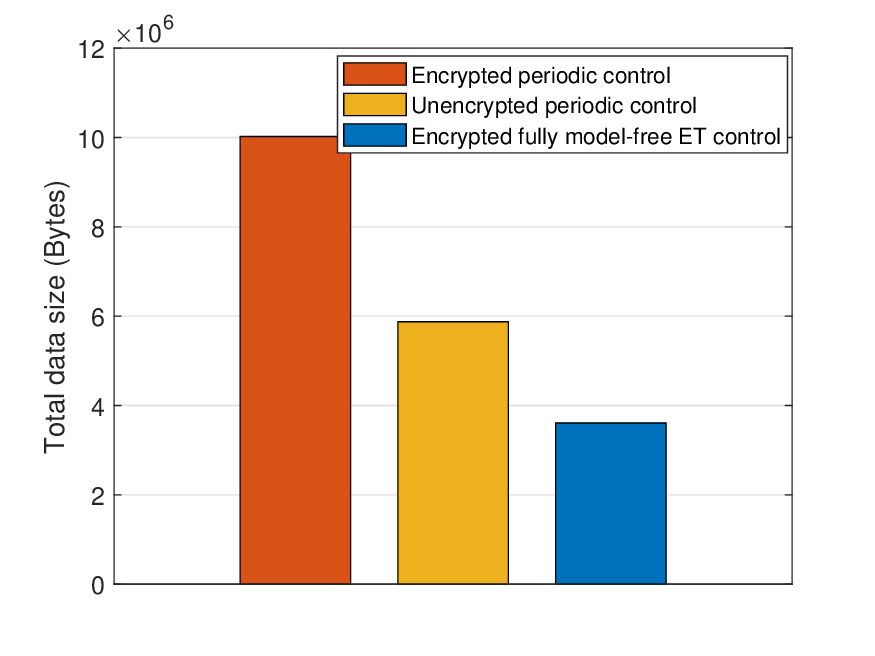}
    \caption{The total communicated data size for encrypted periodic control, unencrypted periodic control, and encrypted fully model-free event-triggered control schemes. Note that ET control denotes event-triggered control.}
    \label{size}
\end{figure}

In Fig. \ref{size}, we present a comparison of the total data size communicated between the cloud and HVAC system over time for three different control schemes: unencrypted periodic neural network based control, encrypted periodic neural network based control, and the proposed encrypted fully model-free event-triggered control with $\beta=6.9$. The data indicates that encrypted periodic control results in a nearly 2 times increased communicated data size compared to unencrypted periodic control. This increase is particularly problematic for wireless sensors, as their communication units are the primary source of power consumption \cite{yadav2016review}. Furthermore, the data size for the proposed encrypted fully model-free event-triggered control is reduced to about 36\% of that for the encrypted periodic control, demonstrating a substantial reduction in both data size and power consumption for sensors.

In terms of computation time, the unencrypted model-free controller requires $0.006$ seconds per time-step, whereas encrypted MPC-based control with $5$ iterations, takes $5.74$ seconds. This indicates that encrypted control demands significantly more computational resources than unencrypted control. However, the proposed encrypted fully model-free control approach requires $2.18$ seconds to compute the control input, which is more than $50\%$ less than that under the encrypted MPC based control. Moreover, the proposed event-triggered communication reduces the required computational resources by $64\%$ as the control input is only updated at triggering instances, thereby reducing the computational resources needed for encrypted control. Thus, the proposed fully model-free framework significantly reduced computation time since the model-free controller does not need to send back encrypted control input many times.

\section{Conclusion}
In this paper, we developed an encrypted fully model-free event-triggered control framework for protecting the sensitive information of building. We first introduced the design of an encrypted model-free controller. This model-free controller was driven by data. Then we applied FHE to the model-free controller. The sensor measurements are encrypted so that the potential leakage can be reduced. What's more, we proposed a model-free event-triggering unit to reduce the high computation cost brought by FHE. On the other hand, a model-free event-triggering unit can avoid the requirements of some prior knowledge for building a model-based event-triggering unit. We finally compared the performance of the proposed encrypted fully model-free HVAC control framework with an advanced encrypted model-based HVAC control framework using the TRNSYS simulator.  

\appendix
\section{A Brief Description of the CKKS Encryption Scheme}\label{App: CKKS}
The CKKS encryption scheme facilitates homomorphic encryption, enabling arithmetic operations on encrypted data, which is critical for privacy-preserving computations. Here, we outline the key components and processes of CKKS.
\subsection{Key Generation}
CKKS involves the creation of:
\begin{itemize}
    \item \textbf{Public Key ($pk$)}: Used for encryption.
    \item \textbf{Secret Key ($sk$)}: Used for decryption.
    \item \textbf{Relinearization Keys}: Utilized to maintain manageable ciphertext size after multiplication. See \cite{fan2012somewhat} for more details on the linearization step.
\end{itemize}
\subsection{Encryption}
The encryption process consists of two main steps:
\begin{itemize}
    \item \textbf{Encoding}: Real numbers are transformed into plaintext polynomials.
    \item \textbf{Encryption}: These plaintext polynomials are encrypted using the public key.
\end{itemize}

Consider the plaintexts $a$ and $b$, the encryption yield ciphertexts $c_a = \enc{a,pk}$ and $c_b = \enc{b,pk}$.

\subsection{Homomorphic Operations}
CKKS supports two primary operations:
\begin{itemize}
    \item \textbf{Addition}: Ciphertexts can be directly added: \\
    $\enc{a+b} = c_a + c_b$
    \item \textbf{Multiplication}: Ciphertexts are multiplied and then linearized to control growth in ciphertext size:\\
    $\enc{ab}=Relin(c_a \times c_b)$
\end{itemize}

\subsection{Decryption}
Decryption is performed using the secret key to convert the ciphertext back to plaintext:

$a+b = \dec{c_a+c_b,sk}$ \quad $ab = \dec{Relin(c_a \times c_b),sk}$

The CKKS scheme is particularly powerful for applications requiring privacy-preserving computations, such as secure machine learning. By allowing operations on encrypted data, CKKS ensures data confidentiality while supporting efficient and approximate arithmetic, which is suitable for practical implementations.

\printcredits

\bibliographystyle{model1-num-names}

\bibliography{cas-refs}


\end{document}